\documentclass[3p]{elsarticle}

\usepackage{amssymb}
\usepackage{graphics}
\usepackage[colorlinks,breaklinks]{hyperref}
\usepackage{hyperref}
\usepackage[utf8]{inputenc}
\usepackage{color}

\journal{NIM A}

\begin{document}

\begin{frontmatter}

\title{Data Acquisition Architecture and Online Processing System\\ for the HAWC gamma-ray observatory}

\author[University of Utah]{A.~U.~Abeysekara}
\author[IF-UNAM]{R.~Alfaro}
\author[UNACH]{C.~Alvarez}
\author[UMSNH]{J.~D.~Álvarez}
\author[UNACH]{R.~Arceo}
\author[UMSNH]{J.~C.~Arteaga-Velázquez}
\author[MTU]{H.~A.~Ayala~Solares}
\author[University of Utah]{A.~S.~Barber}
\author[UMD]{B.~M.~Baughman}
\author[UPP]{N.~Bautista-Elivar}
% ...

\cortext[corresponding]{Corresponding author; currently at CPPM / Aix Marseille Univ., 163 avenue de Luminy, 13288 Marseille, France}

\address[University of Utah]{Department of Physics and Astronomy, University of Utah, Salt Lake City, UT, USA}
\address[IF-UNAM]{Instituto de Física, Universidad Nacional Autónoma de México, Mexico D.F., Mexico}
\address[UNACH]{Universidad Autónoma de Chiapas, Tuxtla Gutiérrez, Chiapas, México}
\address[UMSNH]{Instituto de Física y Matemáticas, Universidad Michoacana de San Nicolás de Hidalgo, Morelia, Michoacan, Mexico}
\address[MTU]{Department of Physics, Michigan Technological University, Houghton, MI, USA}
\address[UMD]{Department of Physics, University of Maryland, College Park, MD, USA}
\address[UPP]{Universidad Politecnica de Pachuca, Pachuca, Hidalgo, Mexico}
% ...
\address[IGeof-UNAM]{Instituto de Geofísica, Universidad Nacional Autónoma de México, Mexico D.F., Mexico}
% ...
\address[IA-UNAM]{Instituto de Astronomía, Universidad Nacional Autónoma de México, Mexico D.F., Mexico}
% ...
\address[ICN-UNAM]{Instituto de Ciencias Nucleares, Universidad Nacional Autónoma de México, Mexico D.F., Mexico}
% ...

\begin{abstract}
The High Altitude Water Cherenkov observatory (HAWC) is an air shower array devised for TeV gamma-ray astronomy.
HAWC is located at an altitude of 4100 m a.s.l. in Sierra Negra, Mexico.
HAWC consists of 300 Water Cherenkov Detectors, each instrumented with 4 photomultiplier tubes (PMTs).
HAWC re-uses the Front-End Boards from the Milagro experiment to receive the PMT signals.
These boards are used in combination with Time to Digital Converters (TDCs) to record the time and the amount of light in each PMT hit (light flash).
A set of VME TDC modules (128 channels each) is operated in a continuous (dead time free) mode.
The TDCs are read out via the VME bus by Single-Board Computers (SBCs), which in turn are connected to a gigabit Ethernet network.
The complete system produces $\approx$\,500 MB/s of raw data.
A high-throughput data processing system has been designed and built to enable real-time data analysis.
The system relies on off-the-shelf hardware components, an open-source software technology for data transfers (ZeroMQ) and a custom software framework for data analysis (AERIE).
Multiple trigger and reconstruction algorithms can be combined and run on blocks of data in a parallel fashion,
producing a set of output data streams
which can be analyzed in real time with minimal latency ($<$ 5 s).
This paper provides an overview of the hardware set-up and an in-depth description of the software design,
covering both the TDC data acquisition system and the real-time data processing system.
The performance of these systems is also discussed.

\end{abstract}


\end{frontmatter}

\twocolumn
\sloppy

\section{Introduction}
\label{sec:intro}
\label{sec:hawc}

The HAWC gamma-ray observatory is an air shower array located on the north slope of Volcán Sierra Negra in central Mexico, at an altitude of 4100~m a.s.l. \cite{HAWC,HAWCSteadySourceSensitivity}.
The experiment is optimized for the detection of gamma rays in the energy range between 100 GeV and 100 TeV.
The HAWC science program includes topics in gamma-ray astronomy, cosmic ray physics, Solar physics and fundamental physics. 
HAWC comprises 300 Water Cherenkov Detectors (WCDs), each holding $\approx$ 200,000 liters of purified water
viewed by four photomultiplier tubes (PMTs) anchored to the bottom.
Three of the four PMTs are 8'' Hamamatsu R5912, these are arranged in an equilateral triangle of side length 3.2~m.
The fourth PMT, positioned in the center, is a high quantum-efficiency 10'' Hamamatsu R7081.
The PMTs observe the Cherenkov light flashes produced by charged particles of the air showers passing through the WCDs.
The array occupies an area of about 170 m in diameter (22,000~m$^2$).
A building situated in the center of the array hosts the front-end electronics, computer farm, and other systems necessary for the functioning of the experiment.

The PMTs are connected via RG-59 cables to the Front-End Boards (FEBs) \cite{Milagrito} reused from the Milagro experiment \cite{Milagro}.
The FEBs are used in combination with Time to Digital Converters (TDCs)
to record the time and the amount of light in each PMT hit exceeding a single photon threshold ($\sim$ 1/4 photoelectrons).
The TDC module chosen for use in HAWC, CAEN V1190A-2eSST, has 128 input channels.
Hence ten TDC modules are necessary for the complete experiment.
The TDCs are synchronized using a common reference clock signal supplied to all TDC modules (40 MHz).
The TDCs are read out by Single-Board Computers (SBCs) using the Versa Module Europa (VME) bus.
The SBCs in turn are connected via a gigabit Ethernet network to the computer farm which processes the data in real time and records the filtered data on disk.
The timing and charge calibration of the PMTs is accomplished using a laser calibration system,
which includes a 45 $\mu$J pulsed laser and a network of optical splitters, fiber optic
switches, and fiber optic cables which distribute the laser light to all WCDs and, via light diffusers, to PMTs \cite{HawcCalibICRC2015}.

The TDCs can by design operate without dead time.
The dead-time free acquisition has given rise to the possibility to operate the HAWC detector without a hardware trigger. 
This scheme, known as the ``software trigger'' scenario, implies that all digitized PMT hits are transmitted to a computer system
and all data filtering is performed in software.
This scheme eliminates the need to develop, operate and maintain a dedicated hardware trigger system which would need to process signals from 1200 PMTs.
It also facilitates implementation of new trigger designs.
The ``software trigger'' approach has been successfully adopted for HAWC.
The TDC readout has been optimized for maximum throughput,
and a dedicated online data processing system has been developed to enable real-time data analysis.
The TDC data acquisition system (DAQ) produces $\approx$\,500 MB/s of raw data
which are reduced to $\approx$\,20~MB/s by the online processing system.
The online processing system uses off-the-shelf computer and network hardware.
All HAWC-specific functions are implemented in software applications,
which includes Data Queues, Reconstruction Clients, Event Sorters and Analysis Clients.
The system uses ZeroMQ \cite{ZeroMQ} to manage data transfers between software components.
ZeroMQ was selected for its simplicity, speed and flexibility.

This paper describes the design of the TDC data acquisition and online processing systems,
with an emphasis on the software architecture of the online processing system.
The paper is organized as follows.
The requirements imposed on the TDC DAQ and the online processing system design, as well as hardware constraints, are explained in section~\ref{sec:requirements}.
Section~\ref{sec:platform} explains the choice of the software platform.
Section~\ref{sec:design} describes the architecture of the developed system and roles of its components.
Section~\ref{sec:performance} discusses the system performance and hardware limitations
and shares the experience from its operation.
Conclusions are summarized in section~\ref{sec:conclusion}.

\section{Performance requirements}
\label{sec:requirements}
The system responsible for the acquisition and online processing of HAWC data must
keep up with the data rate from the continuously operating detector.
Given the PMT count rates observed in HAWC WCDs (25 kHz for the 3 peripheral PMTs and 45~kHz for the 1 central PMT), 
a data rate of nearly 500~MB/s is produced by the complete system prior to data reduction.
The TDC DAQ must continuously acquire this data stream,
which implies transferring $\approx$\,50 MB/s over a VME back plane from each TDC to its SBC.
The ensemble of TDC/SBC pairs should operate synchronously for the full duration of a physics run ($\sim$\,24~hr).
This requires appropriately configured trigger and synchronization signals, as well as a reliable low-latency readout software.
The limited size of the TDC output buffer (128 kB in total for the 128-channel TDC module)
imposes a strong constraint on the TDC readout latency of about 1 ms or better.

The online farm must be able to receive and process the data in real time with a minimal latency.
It needs to reduce the 500~MB/s raw input data to $\approx$\,20~MB/s, the design target for long-term storage of data.
This design target is a compromise between the needs of the physics analyses and the cost of the storage space.
It allows to accommodate a 25~kHz rate of triggered air shower events, including the PMT hit data.
The processing may include trigger algorithms, reconstruction of shower core and direction, background suppression, etc.
Due to the high data rate this processing stage must be parallelized.
The results of the processing need to be collected from across the CPU farm, sorted and saved to disk.
In order to minimize the disk load and reduce latencies it is also desirable to support the real-time analysis of the resulting data stream,
bypassing the disk storage.
The analysis may include searches for gamma-ray bursts (GRBs) and other transients,
producing sky maps, data quality monitoring, and specialized triggers for exotic particles.

Different types of analyses may be implemented as independent applications that each receive a copy of the same data.
The latency requirements are mainly driven by the GRB searches, which are used to trigger multi-wavelength follow-up campaigns \cite{HAWC}.
Given the characteristic time scale of short GRBs ($\sim$ 1 s), latencies $\lesssim$\,1 s are desirable.
Minimizing the alert latency will maximize the chances of a successful follow-up and could allow for a more detailed measurement of the GRB light curve,
e.g. using a narrow field-of-view instrument such as VERITAS \cite{Veritas}.
Early delivery of analysis results may also enhance the capabilities of simultaneous multi-wavelength searches, performed,
e.g., in the framework of the Astrophysical Multimessenger Observatory Network (AMON)~\cite{AMON}.

The data rates are well within the limits of modern off-the-shelf network hardware.
However, collecting all of the raw data at a single computing node
would require a server with CPU power exceeding that of most presently available off-the-shelf products,
and a network bandwidth in excess of 4 Gbps,
and so the architecture of the data network should include direct connections between the data sources (SBCs) and the processing clients, a ``many-to-many'' data transfer scheme.
The total computational power required for the real time processing of HAWC data was estimated to be $<$ 200 CPU cores
when using typical modern CPUs similar to AMD Opteron$^{TM}$ 6344.
This is provided by a set of four 48-core servers, each using two bonded 1~Gbps network interfaces for receiving raw data.

\section{Software platform}
\label{sec:platform}
HAWC uses a modular C++ software framework called AERIE (Analysis and Event Reconstruction Integrated Environment) for all of its reconstruction and analysis software.
It is a modular system providing a consistent interface for the development of reconstruction components, written for HAWC, but sharing a number of concepts from the IceTray framework for IceCube \cite{IceTray}.
The AERIE framework provides a convenient platform for building data processing applications,
but it lacked dedicated tools for passing the data across the network.
Complementing AERIE with a simple data transfer library provides a straightforward way towards a complete data processing network.
We evaluated several software technologies used inside and outside the scientific community before ultimately adopting the ZeroMQ software library \cite{ZeroMQ}.
ZeroMQ is a lightweight and actively developed open-source platform for distributed computing.
Of several technologies evaluated, we found that ZeroMQ best supports the needs of the HAWC data transfer library, which require:

\begin{enumerate}[(a)]
  \item A simple and efficient method to transfer data over the network, hiding the complexity of (proper handling of) TCP/IP sockets.
  \item Support for multiple connections between elements organized in a many-to-many fashion (i.e., no ``central broker'').
  \item Scalability to the network size and data rates expected in HAWC.
  \item Modest CPU and memory consumption.
  \item Compatibility with the AERIE framework (i.e., C/C++ bindings) and the computing environment (OS version, compiler, etc.) used in the online system.
  \item Ease of use and maintainability, with ample freedom to design data formats and code structure.
  \item Availability free of charge, preferably as an open source package.
\end{enumerate}

ZeroMQ was selected for the HAWC online processing system because it meets all of these requirements.
Data packets are treated by ZeroMQ as byte arrays, leaving the data format definition to the user.
It allows multiple connections per socket, which simplifies connection logic in a many-to-many data network.
Its ease of use is particularly attractive for an experiment with a relatively small number of collaborators such as HAWC.

The raw data transmission uses a custom binary format based on the CAEN TDC data format.
Thus the raw data blocks can be sent across the network using ZeroMQ ``as is.''
However, the results of the data processing by the online farm need to be serialized to obtain a byte array suitable for transmission using ZeroMQ.
This is accomplished using the XCDF (eXplicitly Compacted Data Format) library \cite{XCDF},
which is employed as the main data format in HAWC.

\section{Overall design, hardware and software components}
\label{sec:design}

\begin{figure*}
  \centering
    \includegraphics[width=0.88\linewidth]{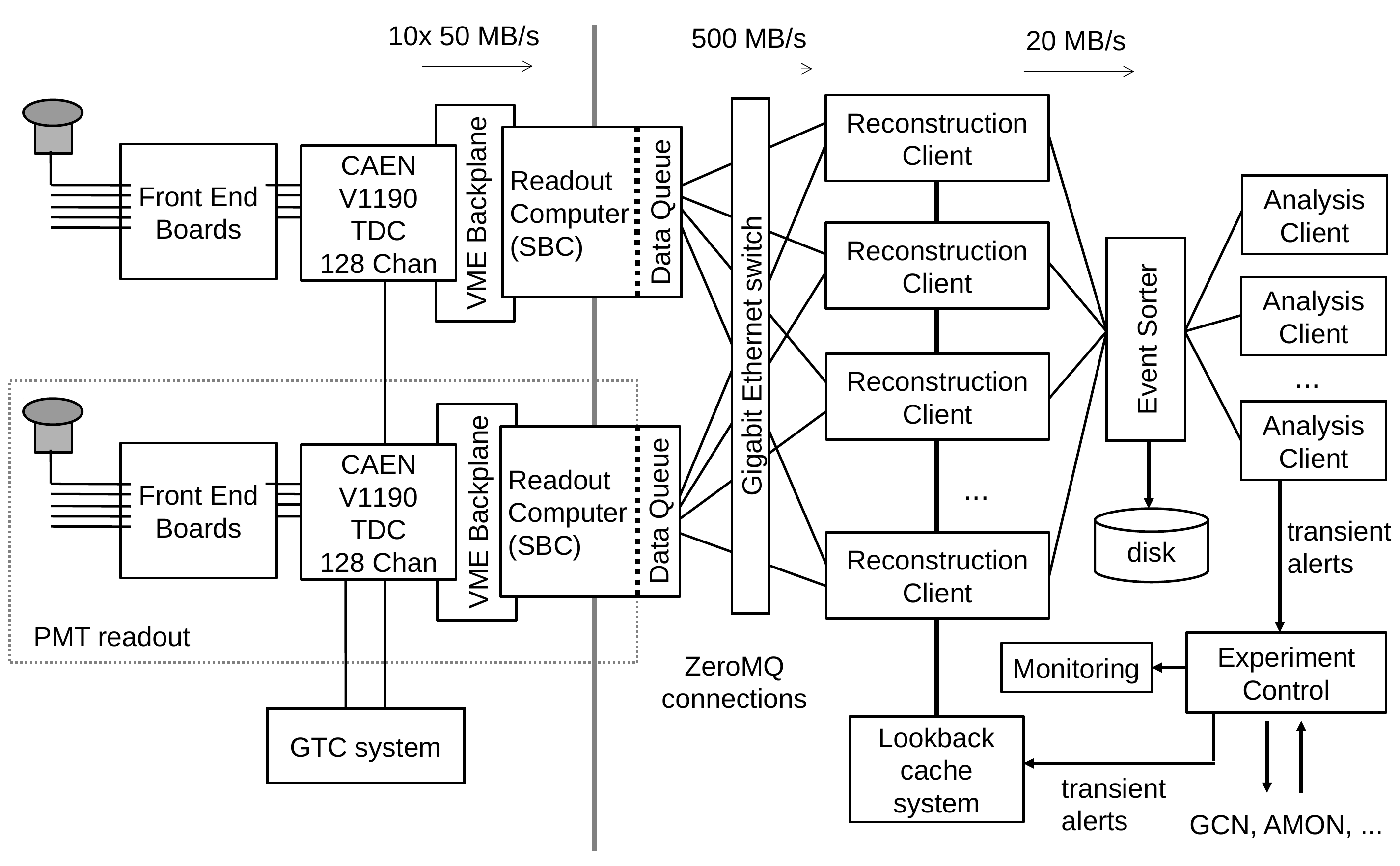}
  \caption{Schematic overview of the HAWC data acquisition and online processing system.
  To the left of the vertical gray line are electronics components of the data acquisition system.
  Each Front-End Board (FEB) receives signals from 16 PMTs.
  Output signals from a group of eight FEBs are digitized by a TDC module,
  which in turn is read out be a dedicated readout computer (SBC).
  In total, ten TDCs and ten SBCs are used to read out 1200 PMTs.
  On the right-hand side of the figure are the software components of the online processing system.
  The network switch connecting the readout computers with the reconstruction farm is also shown.
  Data connections between components are indicated by thin lines.
  Service connections between the Experiment Control and the controlled components are not shown.
  Different software components may run on different machines.
  For further details see text.
  }
  \label{fig:layout}
\end{figure*}

The design of the TDC data acquisition and online processing systems is illustrated in Fig.~\ref{fig:layout}.
The PMT signals processed by the Front-End Boards are digitized by the TDCs, which receive synchronization signals from the GPS Timing and Control (GTC) system (see below).
The data are retrieved from the TDCs by the readout computers (SBCs) using a custom software application called ``Readout Process``.
The data are pushed by the Readout Process to a ``Data Queue`` application and then passed to the reconstruction farm (via a network switch).
The reconstruction farm is comprised of several computers running identical copies of an application called ``Reconstruction Client.''
The Reconstruction Client is responsible for assembling the data from all TDC fragments and applying triggers and reconstruction.
The results of the data processing by the Reconstruction Clients are collected by one or more copies of the Event Sorter application,
which saves the data to disk and forwards it to Online Analysis Clients.
The status of all components is monitored and controlled by a custom software tool called Experiment Control.
The communication between all software applications employs ZeroMQ sockets.
The role of each component is explained in detail below.
The overall design and naming of the components was inspired in part by the ANTARES data acquisition system \cite{AntaresDAQ}.

\subsection{Front-end boards}
The Milagro Front-End Boards (FEBs) \cite{Milagrito} receive PMT signals via RG-59 cables,
which are of equal length for all PMTs (610~ft or $\approx$\,186~m) in order to preserve the signal timing.
The FEBs have 16 inputs.
The analog and digital parts of the FEBs are physically implemented as two separate boards which are installed in modified Versa Module Europa (VME) cages and connected to each other through the cages' backplanes.
On the analogue part of the FEB, the PMT signals are amplified and integrated in a capacitor circuit with a characteristic discharge time of 100 ns, broader than the typical width of an air shower signal ($\sim$\,10~ns).
For each input, the resulting signal amplitude is compared to two pre-set thresholds,
forming two digital signals which are then multiplexed by the digital FEB.
An Emitter Coupled Logic (ECL) signal with two or four edge transitions is produced
depending whether one or both thresholds were crossed by the PMT pulse.
In the case of a two-edge event, the rising and falling edges simply correspond to the crossings of the low discriminator threshold ($\sim$~1/4~photoelectrons).
The two additional edges present in a four-edge event have the opposite polarity and correspond to the high threshold ($\sim$~4~photoelectrons) crossings
delayed by 25 ns (see \cite{Milagrito} for details).
Both the time and amplitude of the PMT pulse can thus be extracted in later analysis from the timing of the edges. 
The analog FEBs are also responsible for generating and distributing the high voltage to the PMTs.

\subsection{TDCs}
The HAWC data acquisition system employs VME Time to Digital Converters (TDCs) to digitize the two and four-edge events from the front-end boards. 
The TDCs from CAEN (V1190A-2eSST) are based on 4 High Performance TDC chips \cite{HPTDC} and designed to operate free of dead time.
All TDC modules operate strictly in parallel. Each acquires data from up to 128 PMTs
and receives the same trigger and synchronization (``external clock'') signals (see sect.~\ref{sect:gtc}).
Ten TDC modules are necessary for the complete experiment.
The TDCs are operated with a 98 ps resolution for the least significant bit,
using 8 bits to represent times between consecutive beats of the 40 MHz clock.

Following the ``software trigger'' paradigm, the TDCs acquire and transmit all of the TDC edges to a computer farm.
Thus, the TDC is used as a continuous recorder of pulse edge timing rather than the usual V1190 application - the triggered event readout.

For each trigger, the TDC module records the times of the edges from each input channel during a pre-set time window that includes the trigger time.
The set of TDC edge times, together with the CAEN auto-generated header data for each TDC chip and TDC trigger (including error words and time information and word counts), form a {\it TDC event}.
A periodic trigger every 25 $\mu$s is used to control the TDC event generation, and the TDC internal readout time window is set equal to the trigger period plus a 1 $\mu$s overlap to avoid dropping hits on the edge of the time window.
This leads to a data rate of $\approx$\,350 kB/s per PMT, or $\approx$\,45 MB/s per TDC module.

The sequential trigger number, as counted by a given TDC module starting from the most recent run start, is included in the data stream.
Thanks to the GTC system, all TDCs are synchronously reset before a run start and receive synchronous triggers during the run,
the ensemble of TDCs are expected to remain in sync at all times.
A timing word, counting the external clock frequency (40 MHz) is also included in the data stream, allowing for an independent cross-check of the synchronization.

A set of 32 input lines on one of the TDCs is reserved for use with the GTC system (Sect.~\ref{sect:gtc}) and calibration signals.
These input signals are treated by the TDC DAQ in exact same way as normal signals from the Front End Boards.

\subsection{Single-board computers}
Each of the five VME back planes in each VME crate (Wiener 6023x610) hosts a TDC module and dedicated single-board computer (SBC) from General Electric (XVB602-13240010) running CentOS.
Thus every TDC-SBC pair uses an independent VME backplane.
The SBC receives data via a CAEN implementation of the 2eSST VME dual edge transfer protocol from the TDC and after minimal processing passes the TDC data fragment downstream via Gigabit Ethernet.
This process is controlled by a dedicated readout software running on the SBC.
Thus the SBC acts as a bridge between the DAQ hardware and the online processing farm. 

\subsection{Readout Process}
The data are retrieved from each TDC module by the Readout Process.
running on a SBC (single-board computer) that shares the VME backplane with the TDC.
The readout makes use of the TUNDRA chipset of the SBC that allows a direct memory access (DMA) transfer of data that is ready to be transferred from the TDC output buffer.
Limited by the clock speed of the TDC VME chip, 
the 2eSST transfer speed on the backplane has a theoretical maximum of 100 MB/s but under run conditions is about 60 MB/s, varying slightly from one TDC/SBC pair to another.

The CAEN firmware sets a ``data ready'' register when a block of 25 TDC events ($\approx$\,625 $\mu$s), on average corresponding to 1/3 of the size of the TDC output buffer, is ready.
The SBC polls the data-ready register and initiates the 2eSST DMA transfer.
The Readout Process transfers the data blocks from the DMA buffer immediately after they become available,
and arranges them into bigger blocks (without ever opening the actual data content) suitable for transmission via Ethernet.
These larger blocks consist of 42 TDC blocks and contain a total of 1050 TDC readout windows (25+1 $\mu$s each).
These blocks are numbered sequentially starting from the most recent run start.
Two of the 42 TDC blocks are redundant with the following block in order to implement tolerance for rare CAEN firmware glitches.
Thus each block is guaranteed to contain a range of triggers (a 25 ms interval) that is defined by the block number.
This ensures that a complete air shower event can be constructed based on data blocks carrying the same block number.
This redundancy is also used in the Reconstruction Clients to eliminate dead time at the block boundaries.
Blocks which are ready for transmission are pushed to the Data Queue running on the same SBC using the ZeroMQ protocol.
In order to ensure a sufficiently low latency of the TDC data polling, the Data Queue and Readout Process threads were locked to different CPU cores
(using the CPU affinity settings) and their priorities were set to high values 
(a set-up mimicking the behavior of a real-time operating system).

\subsection{GTC system}
\label{sect:gtc}
The HAWC GPS Timing and Control (GTC) system is a custom hardware system which has a number of responsibilities related to timing, triggering and control.
In particular, it provides a common clock signal (40 MHz) and a periodic trigger signal (40 kHz), which are fanned out to all TDC modules.
The starting/stopping of the data acquisition is performed by enabling/disabling the trigger signal.
The system is also responsible for inhibiting triggers when data acquisition is to be turned off,
and generating signals that can initiate a hardware clear and reset for the timing and event counters of the TDCs, in order to keep all TDCs synchronized.
Special care is taken to ensure simultaneous delivery of these signals to all TDCs.
The GTC system also provides a 28-bit global timestamp to the TDC DAQ.
The system internally connects to a GPS receiver and re-formats the current time into a sequence of edges on 28 output lines.
In this scheme, a 1~$\mu$s pulse is used to denote a logic zero bit, and a 2~$\mu$s pulse for a logic one bit.
These 28 lines are fed into a TDC and read out along with the event data.
The data are used by the Reconstruction Clients to timestamp physics events.
A complete description of the HAWC GTC system will be published elsewhere \cite{GTCpaper}.

\subsection{Data Queue}
The Data Queue is a software application that caches the data retrieved from the TDC by the readout process
and serves it up when requested by the Reconstruction Clients (see Sect.~\ref{sect:recoclient} and Fig.~\ref{fig:layout}).
One Data Queue is used for each readout process.
Both processes run on the SBC.
One of the Data Queues is designated as the ``master'' for organizing the assignment of the data to a Reconstruction Client.

\subsection{Reconstruction Client}
\label{sect:recoclient}
The Reconstruction Client is a software application which processes the raw TDC data,
{bf
searching for air-shower events and determining the direction and event parameters of the identified showers.
It also plays a key role in organizing the data flow in the online processing system.
Whenever ready,} Reconstruction Clients poll the master Data Queue for the next block of data.
While the block size (25 ms in the initial implementation) is chosen to reduce the effect of any overhead in initiating a transfer, the actual air shower event size is much smaller ($\approx$ 2 $\mu$s).
Upon receiving the next block to process, the clients poll the remaining Data Queues for the data from the other TDCs acquired during that block.
As a result of this scheme, each block of raw data is only transmitted once over the network.
The set of concurrent data blocks from all TDCs covering a 25 ms period is called a ``time slice.''
After receiving a complete time slice, i.e. the correct blocks of data from each SBC/TDC, the Reconstruction Client decodes the data
and applies physics trigger, calibration, and reconstruction algorithms.

The processing of each time slice uses only the information contained in that time slice.
Hence multiple copies of the Reconstruction Client can work in parallel without the need to communicate with each other.
The number of clients assigned to a given physical machine is chosen according to the available resources.
The utilization of any reconstruction machine decreases in a natural way when additional Reconstruction Clients are added on an external machine, 
allowing the system to scale in a robust way.
The first and last sub-block of each time slice are redundant, appearing in the previous and next time slice, respectively.
This redundancy eliminates deadtime from separating physics events across time slice boundaries.

The Reconstruction Client is implemented as a standard AERIE application and is composed of software components termed ``modules.''
Any algorithms developed within the AERIE framework can therefore be used in the online system without changes.
The data decoding, trigger, calibration and reconstruction algorithms operate in a chain,
with each module selectively using the output from the preceding modules.
The only difference between the online Reconstruction Client and the corresponding offline application
is the use of different input and output modules (ZeroMQ socket I/O vs. disk I/O).
This design provides numerous advantages over a more traditional standalone DAQ application:

\begin{enumerate}
\item High-level event processing (event reconstruction and filtering) is incorporated directly with event triggering and low-level data processing, eliminating any interface between these two systems.

\item Use of modular pieces allows rapid, independent development and incorporation of trigger algorithms.

\item Development of the data acquisition chain can be done with a simple AERIE application using existing raw TDC data files simply by substituting the ZeroMQ socket data reader with a file reader.

\item Experience with AERIE in the collaboration is utilized, resulting in a larger pool of scientists capable of contributing to trigger and online reconstruction development.

\item The same processing and event triggering algorithms can be used for online data acquisition as well as offline data analysis and simulations.
\end{enumerate}

\begin{figure}
  \centering
    \includegraphics[width=0.98\linewidth]{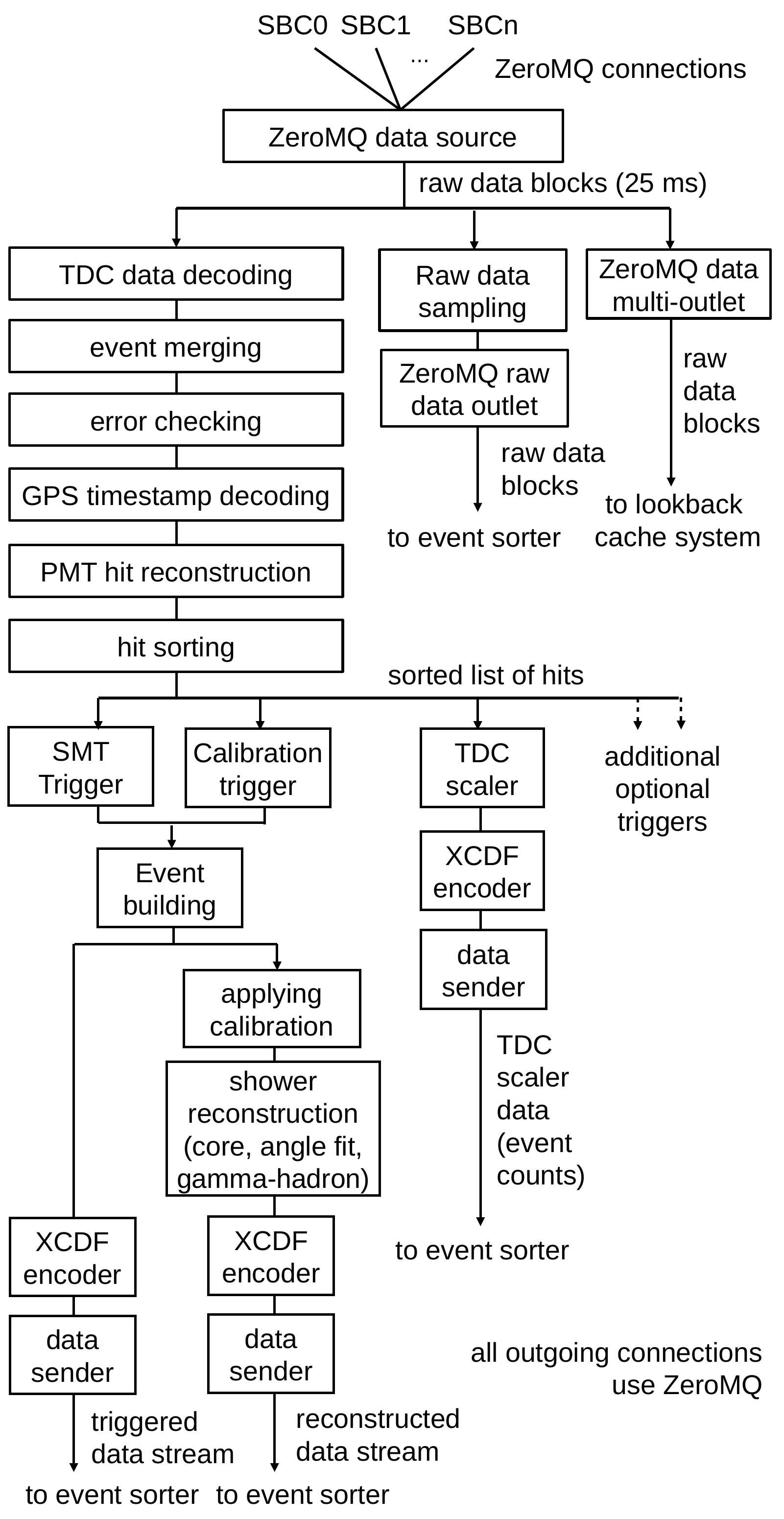}
  \caption{Schematic overview of the HAWC Reconstruction Client.
  The software modules included in the Reconstruction Client are shown by rectangles.
  The arrows indicate the direction of the data flow.
  See text for details.
  }
  \label{fig:recoclient}
\end{figure}

Figure~\ref{fig:recoclient} gives an overview of the software modules included in the Reconstruction Client and the data flow in the module chain.
For each time slice, all relevant software modules are executed, starting from the Data Source and ending with the Data Senders.
The Data Source is a service module responsible for polling the Data Queues as explained above.
The Data Senders are tasked with sending the output data to the Event Sorters (see Sect.~\ref{sect:eventsorter}).
The functionality of the other modules is explained in the following sections.

\subsubsection{Initial TDC Data Processing}
The initial data processing is performed by a chain of six software modules, shown in Fig.~\ref{fig:recoclient}.
Each time slice is first broken down into data from the constituent TDCs, and then further broken down into the 26~$\mu$s readout windows.
For each readout window, data from each TDC is merged and checked to ensure the TDCs are properly synchronized.
One merged readout window represents a complete 26~$\mu$s readout of the $\approx$1200 detector channels.
Next, the GPS timestamp embedded in the TDC data is decoded, providing an absolute time for the start of the readout window.
A timestamp retrieved from a local NTP server is also recorded, providing an alternative global timestamp with a $\sim$50 $\mu$s resolution.
After that, the TDC data from PMT channels is decoded into an absolute time for each edge.
The time is represented by a 64-bit unsigned integer in units of the TDC resolution (98 ps), with zero corresponding to the beginning of the GPS epoch (Jan 6, 1980, 00:00:00 UTC) \cite{GPS}.
The time assigned to an edge is the GPS time corresponding to the beginning of the readout plus the offset of the edge relative to the beginning. 
Using this global time, duplicate edges that correspond to the 1 $\mu$s overlap between consecutive readout windows are identified and removed.
Rising and falling TDC edges are then grouped together into PMT pulses (hits).
The hits in the time slice are sorted in time, with the hit time corresponding to the time of the leading edge.

\subsubsection{Triggering and Event Building}

The time-sorted sequence of hits is used by several software trigger modules to identify interesting events.
The result of any trigger module is a list of times at which the trigger condition was satisfied during the time slice.
The two triggers currently in operation are: 
\begin{enumerate}
\item Calibration events, which are triggered by a dedicated TDC channel connected to the HAWC laser calibration system;
\item a Simple Multiplicity Trigger (SMT), which is the main air-shower physics trigger, requires that $n$ hits are observed within $t$ nanoseconds, and both $n$ and $t$ are configurable options.
With 294 WCDs, HAWC used a multiplicity window of $t$ = 150 ns and a threshold of $n$ = 28 PMTs, resulting in a trigger rate of 25~kHz.
\end{enumerate}
The lists of trigger times from each of the trigger modules are processed by the event builder module.
For each trigger, a configurable time window around the trigger is defined, and the hits within this window are extracted.
In the case that multiple triggers fire within the event time window, the window is enlarged to appropriately contain all the hits corresponding to the trigger.
Events with trigger time within the two redundant sub-blocks at the edges of the time slice are dropped.
This prevents duplication of the same physics events and eliminates dead time at time slice boundaries.
The resulting set of events is placed into a list for further processing and reconstruction and is transmitted to the Event Sorter for archival.

The multiplicity threshold used for the SMT ($n$ = 28) is lower than that used in most physics analyses \cite{HAWC,HAWCSteadySourceSensitivity}.
This ensures that the trigger does not limit the HAWC's capabilities in any important way.
The threshold can be further lowered, if needed,
provided that sufficient computing and data storage resources are available to accommodate the increased data rates.

\subsubsection{Reconstruction}

The events in the time slice are fed one-by-one to the calibration and reconstruction modules
(all modules being pieces of C++ code plugged into the AERIE framework).
Calibration corrects for minor time offsets between channels and slewing (the dependence of the discriminator crossing time on the pulse amplitude)
and converts the TDC time-over-threshold into observed PMT charge.
The calibrated PMT times and charge are used in air shower core and angle reconstruction.
The data are also used to characterize the air shower as a likely gamma-ray or a hadronic shower.
The laser calibration events (flagged by the Calibration trigger) are ignored by the air shower reconstruction.
Reconstruction results are transmitted to the Event Sorter using XCDF as data serialization format.

\subsubsection{Auxiliary data streams}
The Reconstruction Clients also include functionality for monitoring the PMT count rates ("TDC scaler"),
recording samples of raw TDC data (e.g. for purposes of data checks),
and re-directing a copy of the raw data stream to the lookback cache system (Sect.~\ref{sect:lookbackcache}).

\subsection{Event Sorter}\label{sect:eventsorter}
The Event Sorter is a software component which collects blocks of data from the Reconstruction Clients.
Since each block of data corresponds to a different range of times, the data arriving at the Event Sorter is sorted to form the final event stream which is saved to disk.
Sorting is done using the block number, as assigned by the readout software.
In order to eliminate any complex data decoding within the Event Sorter, 
each data packet transmitted by the Reconstruction Clients to the Event Sorter includes the block number as a 32-bit word,
preceding the main block data.
The main part of the data block is saved to disk without decoding, reducing computational overhead.
The Event Sorter starts a new file each time a configurable block count limit is reached ($\sim$ 2 min), defining a ``sub-run.''
Depending on the data format used, at run start the Reconstruction Clients may send a special data block to be used as a file header.
This block is identified using a distinctive value of the block number and is treated in a special way by the Event Sorter.
If such a header block is used, its copy is written at the beginning of each sub-run.
The Event Sorter can also be configured to write a predefined byte sequence (a file trailer) at the end of each sub-run.

An XCDF file contains a file header, a sequence of data blocks, each of which can be interpreted independently, and a file trailer.
The file header can be assembled before the data start flowing and is therefore identical for each of the Reconstruction Clients running in parallel.
The file trailer is intended to store an event lookup table for random access but can be left empty.
The independence of the data blocks allows for the combination of blocks received from different sources in a straightforward way.
This permits the Event Sorter to assemble a valid XCDF file without decoding the data.

The data are also forwarded to various analysis processes that may be connected to the Event Sorter.
In the current implementation four Event Sorters are used to accommodate the following four data streams: triggered data stream (includes PMT hit data),
reconstructed data stream (reconstructed direction, fit quality, etc.), ``TDC scaler'' data (PMT rates), and raw data samples.

\subsection{Analysis Clients}
Analysis Clients are applications that receive and process a copy of a data stream from an Event Sorter.
Typically, each Analysis Client is in charge of a specific type of analysis.
More than one Analysis Client can be used with each Event Sorter.
Analysis Clients should be fast enough to cope with the full data stream they subscribe to.
Examples of Analysis Clients include the production of sky maps and temporal searches for GRBs and AGN flares.

\subsection{Lookback cache}\label{sect:lookbackcache}
The Lookback cache is a distributed system that can receive a copy of the raw data stream from the Reconstruction Clients
and store the full 500 MB/s data in a temporary disk cache.
Each computer that runs Reconstruction Clients is equipped with sufficient number of hard drives to accept the raw data available on that computer
and store it for $\sim$\,24~hr ($\sim$\,10~TB for each of the four servers).
Upon request, portions of the data may be marked as ``useful'' and moved to a permanent disk storage.
This functionality is primarily intended for enhancing the sensitivity to low energy gamma rays from GRBs.
At the time of writing this manuscript the disk cache was only partially implemented, pending a detailed scientific justification.

\subsection{Experiment Control}
Experiment Control is the software system for monitoring and controlling the state of the experiment.
Experiment Control is responsible for obtaining the desired operating settings for the detector and data acquisition software,
recording and communicating those parameters to the components under its control, and starting / stopping runs.
The communication of major experimental errors also goes through Experiment Control.
In the case of a major error, Experiment Control will attempt to automatically restart the run.
An automatic run restart normally takes 3-4 min to complete.
Communication of commands and status messages between the Experiment Control and the controlled components is
implemented using the cJSON library \cite{cJSON} for data serialization.
Experiment Control also manages the communication with external transient notification systems such as
the GRB Coordinates Network (GCN) \cite{GCN}.

\subsection{Monitoring}
The status of the experiment is monitored using a dedicated software package written in Python.
The monitoring data, such as PMT count rates and GTC status, is obtained via Experiment Control or directly from the monitored components using ZeroMQ sockets.
The collected data is stored in a MySQL database at the HAWC site and propagated to remote servers at two different sites
(University of Maryland and National Autonomous University of Mexico) for redundancy.
The monitoring system uses Google Charts API to make plots of monitored quantities.
The plots are made available to the experiment operators in near-real time via a web interface.

\section{System performance}
\label{sec:performance}
\subsection{Operation experience and TDC DAQ performance}
The TDC DAQ and online processing system have been used for data taking in HAWC starting from 2012, when first WCDs were installed,
with only minor changes to the system design in the following years.
The experience from operating these systems has met the expectations.
In particular, it has been shown that each TDC is able to digitize the signals from 128 PMTs in a continuous mode
and the full data stream can be transferred from the TDC to the SBC, meeting both the throughput ($\gtrsim$\,50~MB/s) and readout latency ($\lesssim$\,1~ms) requirements.
The average latency of TDC readout was measured during normal data taking and found to be 0.625 ms.
A set of 10 TDC-SBC pairs can run synchronously for at least 24~hr.
The CPU and memory resources available on the SBC are sufficient for the TDC readout software and Data Queue.

\subsection{Present configuration of the online system}
Presently the HAWC online processing system utilizes four rack-mounted servers,
with 48 CPU cores on each server (four 12-core AMD Opteron$^{TM}$ 6344 CPUs).
The first three servers are used to run Reconstruction Clients (one client per core).
The fourth server accommodates the Event Sorters and Analysis Clients, as well as additional Reconstruction Clients.
The servers and SBCs are connected via a 1~Gbps Ethernet switch (HP 2910-48G) which has
48 ports, 176 Gbps switching fabric, and up to 131 Mpps (millions of packets per second) packet throughput.
This provides a total switching capacity well exceeding the HAWC requirements.

\subsection{Throughput}
Each server has four 1~Gbps interfaces which can be logically ``bonded'' together.
Presently just two bonded interfaces per server are used for receiving raw data.
This theoretically allows a total data rate up to 8~Gbps.
Similarly, the SBCs are configured to each use a bonded pair of 1 Gpbs interfaces.
Network throughput measurements performed in this network configuration suggest that the bonded interfaces provide a throughput of about 0.7~Gbps times the number of subordinate interfaces.
This leaves a performance margin of about a factor of 3 for the SBCs and a factor of $\approx$\,1.5 for the reconstruction farm (assuming a total data rate from the SBCs of 4~Gbps).
This performance margin is sufficient to cope with a failure of one of the servers, and
the software design guarantees that the load will be automatically distributed among the Reconstruction Clients running on the remaining servers.

Our tests show that a point-to-point connection between two ZeroMQ sockets in a 1~Gbps network using 1~MB blocks
can deliver a throughput of $\approx$ 940 Mbit/s, which is very close to the theoretical maximum bandwidth of such a connection.
This suggests that ZeroMQ adds only a marginal overhead to data transfers.
Tests indicate that the use of blocks much smaller than 1~MB tends to reduce the throughput by a significant factor,
but increasing the block size above 1~MB does not affect the throughput.
The use of a ``request-reply'' pattern, as defined by ZeroMQ,
for communication between the Data Queues and Reconstruction Clients
leads to a somewhat lower throughput due to the waiting times associated with that pattern.
This reduction is the largest in the case of a single client and small data blocks.
Using 1 MB data blocks and three or more clients per Data Queue produced a bandwidth reduction of $\approx$ 10\% compared to the simplest one way communication pattern.
As a countermeasure, the Reconstruction Clients implement a procedure for advance requests, meaning that a new data block can be requested while the current block is being processed.

The data produced by the Reconstruction Clients (and delivered to the Event Sorters) make a small contribution to the network traffic ($\approx20$ MB/s).
The lookback cache design implies that the data are stored locally on each server.
Its operation therefore does not generate any network traffic except when data are retrieved from the cache.
It is straightforward to increase the capacity of online processing system, if necessary, by adding more servers subject to the limitations of the network switch.

\subsection{Scalability}
Tests using multiple data sources (up to 10) and multiple destinations (up to 100) demonstrated good scalability of the system: 
in all configurations the measured throughput attained $\ge$\,80\% of the hardware limit.
Tests with even larger number of interconnected elements did not show any evidence of a scalability limit.
Hence the online processing system can be considered fully scalable with regard to the number of data sources and Reconstruction Clients.
Experience from operating the online processing system of the HAWC experiment
confirms the high performance and scalability of the ZeroMQ-based design.
It should be noted however, that the Event Sorter has a limited capacity to accept data and write data to disk, and the use of a single Event Sorter to collect the output from all Reconstruction Clients has a limited scalability.

\subsection{CPU consumption}
The CPU consumption associated with data transfers via ZeroMQ is small (a few \% or less) in most cases,
playing a visible role only for the Data Queue and Event Sorter.
The CPU consumption by the Data Queue during normal operation (45 MB/s), including the ZeroMQ-associated consumption,
corresponds to about 7\% of the CPU resources on the SBC, the rest being available for the DAQ readout process.
The Event Sorter consumes less than 20\% of the resources of one CPU core when operating at the design data rate (20 MB/s).
Unlike the Event Sorter, the online Analysis Clients have to deal with decoding the XCDF data stream.
With the current implementation, using a single thread for data decoding and analysis, the clients are limited to data rates below $\approx$ 40 MB/s.
The CPU consumption by the Reconstruction Clients is completely dominated by the raw data decoding and reconstruction algorithms.

\subsection{Data analysis latency}
\begin{figure}
  \centering
    \includegraphics[width=0.7\linewidth]{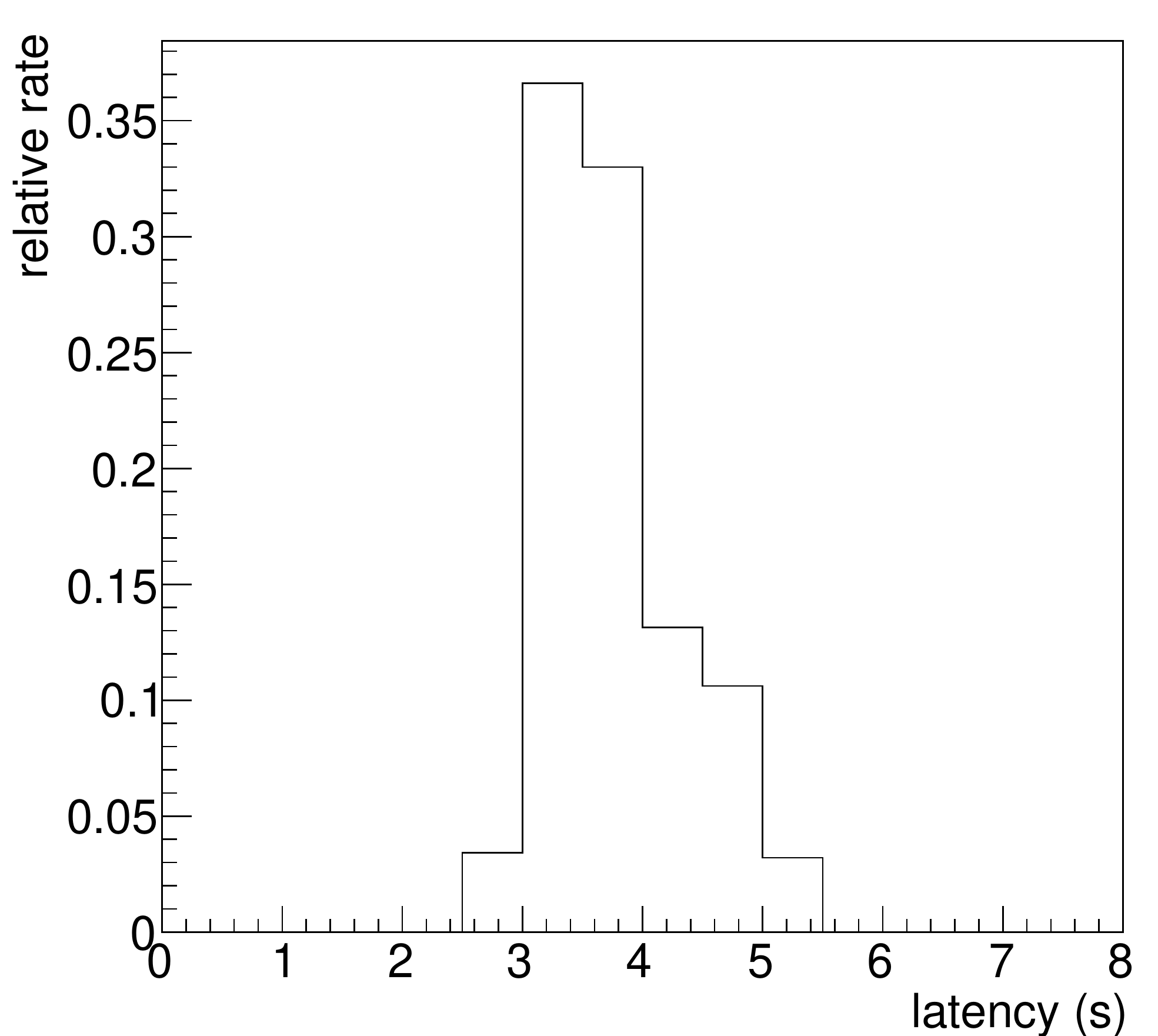}
  \caption{The latency of the online system,
measured from when an air shower event is read out by the SBCs to when it is received by the Analysis Client.
Measurements made during a standard physics run (24~hr) with the complete HAWC array.
  }
  \label{fig:latency}
\end{figure}

The size of the data blocks transmitted from the SBCs to the HAWC Reconstruction Clients is an adjustable parameter.
Increasing the block size tends to reduce the overhead associated with the transmission of data,
but increase the time needed to process the larger data blocks, resulting in a delay in the availability of the output.
HAWC currently uses 25 ms time slices.
This corresponds to $\approx$ 1 MB of data per block for each SBC, which is large enough to guarantee the high throughput.
The processing of the data from the full experiment (300 WCDs) requires $\approx$ 100 CPU cores to run the Reconstruction Clients.
Each Reconstruction Client spends approximately 3 s to process one 25 ms time slice.
The processing time is dominated by the raw data decoding, triggering, calibration and reconstruction algorithms,
with only a small overhead due to the framework.
The processing time may vary slightly ($\sim$ 10\%), depending on the contents of the time slice and the server load.
The availability of the reconstruction results at the output of the Event Sorter is determined by the slowest-to-process time slices.
This currently limits the latency of the online analysis to $\gtrsim$ 3 s.
A measurement of the online analysis latency performed during normal physics data taking with 294 WCDs
yields an average latency of $\approx$ 4 s (Fig.~\ref{fig:latency}).
The latency could be reduced by adjusting the time slice size, if necessary (e.g. for a gamma-ray transient search).

\section{Summary}
\label{sec:conclusion}
The data acquisition system of the HAWC observatory relies on Time-To-Digital converters
to record the time and charge of the PMT signals processed by the Milagro Front-End Boards.
The TDC DAQ system employs ten VME TDC modules, each receiving signals from up to 128 PMTs.
The TDCs are read out by Single-Board Computers via the VME back plane.
The system supports the continuous readout mode, 
where the TDC event generation is controlled by a periodic trigger signal (40~kHz).
The TDC readout proceeds in blocks of 25 events (625 $\mu$s),
controlled by custom software specially optimized for sub-millisecond latency.
The data rate is 45 MB/s per TDC.

The data acquired from the TDCs are processed in real time by a specially designed data processing system.
This system utilizes standard network and computing hardware
and relies on a lightweight open source platform - ZeroMQ - for organizing data transfers.
The software layer of this system consists of three parts: the data collection subsystem, the processing subsystem,
which includes an array of identical processes operating independently on blocks of raw data,
and the online analysis subsystem, intended for high level analysis of the output of the previous stage.
All software components in this system communicate using ZeroMQ sockets.
The system features a scalable design, 
employing direct transfer of data blocks from data sources to processing clients in a many-to-many fashion.
This design, in combination with the use of a high-performance network switch,
enables a high throughput which meets the design requirement for the input rate of raw data of 500 MB/s.
The analysis of this data stream in real time is ensured by $\approx 150$ instances of the data processing client running
on the on-site computer farm,
utilizing fast data encoding/decoding methods (byte arrays or XCDF blocks over ZeroMQ)
and efficient analysis algorithms implemented in C++ code.
All relevant components have been integrated with the HAWC software framework, AERIE.
The reconstruction results are stored using the XCDF format, which is suitable for use in a distributed environment.

The continuous readout mode ensures that every photomultiplier hit is digitized by the TDCs.
Triggering and reconstruction occur entirely in software (the ``software trigger'' scenario).
This approach provides great flexibility in trigger designs, including the possibility to create specialized triggers, e.g. for slow-moving particles such as Q-balls or magnetic monopoles.
It also provides unique opportunities for improving the sensitivity of GRB searches by temporarily reducing the trigger threshold \cite{HAWC}.
The results of the data processing are made available for analysis in real time as soon as they have been collected,
with a latency of $\sim$ 3 s using appropriate settings.
Examples of real-time analysis that are applied to HAWC data include searches for gamma-ray transients, making sky maps, detector status monitoring, etc.
The developed system has been successfully used for HAWC data taking since September 2012,
proving to be stable, fast, scalable and robust.
Since December 2014 the system is continuously handling a 450~MB/s data stream from the fully completed HAWC array (300 WCDs).

\section*{Acknowledgments}
We acknowledge the support from: US National Science Foundation (NSF); US Department of Energy Office of High-Energy Physics; The Laboratory Directed Research and Development (LDRD) program of Los Alamos National Laboratory; Consejo Nacional de Ciencia y Tecnolog\'{\i}a (CONACyT), M\'exico; Red de F\'{\i}sica de Altas Energ\'{\i}as, M\'exico; DGAPA-UNAM, M\'exico; and the University of Wisconsin Alumni Research Foundation.

The Front-End Boards were designed and built at University of California, Santa Cruz.
The Readout Process includes code developed by the Jefferson Lab Data Acquisition Group.


\begin{thebibliography}{}

\bibitem{HAWC} A.~U.~Abeysekara et al., On the sensitivity of the HAWC observatory to gamma-ray bursts,
Astropart. Phys. 35 (2012) 641

\bibitem{HAWCSteadySourceSensitivity} A.~U.~Abeysekara et al., Sensitivity of the high altitude water Cherenkov detector to sources of multi-TeV gamma rays, Astropart. Phys. 50-52 (2013) 26

\bibitem{Milagrito} R. Atkins et al., Milagrito, a TeV air-shower array, NIM~A 449 (2000) 478

\bibitem{Milagro} A. A. Abdo et al., Recent Results from the Milagro Gamma Ray Observatory,
Nucl. Phys. B Proc. Suppl. 151 (101) 2006

\bibitem{HawcCalibICRC2015} H.~A.~Ayala Solares et al., The Calibration System of the HAWC Gamma-Ray Observatory, PoS ICRC2015 (2016) 997, arXiv:1508.04312

\bibitem{ZeroMQ} ZeroMQ documentation,
\url{http://zeromq.org}

\bibitem{Veritas} T. C. Weekes et al., VERITAS: The Very Energetic Radiation Imaging Telescope Array System, Astropart. Phys. 17 (2002) 221

\bibitem{AMON} M. W. E. Smith et al., The Astrophysical Multimessenger Observatory Network (AMON), Astropart. Phys. 45 (2013) 56

\bibitem{IceTray} T. DeYoung, IceTray: a Software Framework for IceCube, 
Proc. Computing in High Energy Physics and Nuclear Physics 2004, CERN-2005-002.463

\bibitem{XCDF} XCDF: The eXplicitly Compacted Data Format,
\url{https://github.com/jimbraun/XCDF}

\bibitem{AntaresDAQ} J. A. Aguilar et al., The data acquisition system for the ANTARES neutrino telescope, NIM~A 570 (2007) 107

\bibitem{HPTDC} J. Christiansen, HPTDC: High Performance Time to Digital Converter, Version 2.0, CERN/EP - MIC, Sep 2001

\bibitem{GTCpaper} A.~U.~Abeysekara, T.~N.~Ukwatta, D.~Edmunds, J.~T.~Linnemann,
GPS Timing and Control System of the HAWC Detector, arXiv:1410.6681
 
\bibitem{GPS} GPS documentation,
\url{http://www.usno.navy.mil/USNO/time/gps/usno-gps-time-transfer}

\bibitem{GCN} S.~D.~Barthelmy et al., GRB Coordinates Network (GCN): A Status Report, Proc. 4th Huntstville GRB Symposium, AIP Conf. Proc. 428 (1999) 99 

\bibitem{cJSON} cJSON documentation,
\url{http://sourceforge.net/projects/cjson}

\end{thebibliography}
\end{document}